\documentclass[letterpaper, 10 pt, conference]{ieeeconf}  % Comment this line out if you need a4paper

\usepackage[pdftex,dvipsnames]{xcolor}  % Coloured text etc.
\usepackage[colorinlistoftodos,prependcaption,textsize=tiny]{todonotes}

\IEEEoverridecommandlockouts                              % This command is only needed if you want to use the \thanks command
\usepackage{amsthm}
\usepackage{graphics} % for pdf, bitmapped graphics files
\usepackage{epsfig} % for postscript graphics files
\usepackage{times} % assumes new font selection scheme installed
\usepackage{amsmath} % assumes amsmath package installed
\usepackage{amssymb}  % assumes amsmath package installed

\usepackage[maxbibnames=99,giveninits=true]{biblatex}
\usepackage{bm}
\usepackage{bbm}
\usepackage{xcolor}
\usepackage{graphicx}
\usepackage{algorithm}
\usepackage[noend]{algpseudocode}
\usepackage{booktabs}
\usepackage{siunitx}% An awesome package for typesetting and manipulation numbers and units.
\usepackage{hyperref}% Used for insert the link

\usepackage{caption}% Better control over caption
\usepackage{lipsum}% Example text
\usepackage{multirow}
\usepackage{diagbox} 
\usepackage{subcaption}
\newtheorem{definition}{\bf Definition}

\newtheorem{theorem}{\bf Theorem}
\newtheorem{lemma}{\bf Lemma}

\newtheorem{remark}{\bf Remark}
\newtheorem{proposition}{\bf Proposition}

\begin{document}
\title{\LARGE \bf
Entropic Perspective on Risk and Robustness in Network of Platoon of vehicles: From tail averages to Exponential bounds
}

\title{\LARGE \bf
Entropic Value-at-Risk for Inter-Vehicle Collision in Platoons: 
Network- and Delay-Induced Bounds on Risk Due to Extreme Events
}

\author{Vivek Pandey, and Nader Motee 
\thanks{
 V. Pandey, and N. Motee are with the Department of Mechanical Engineering and Mechanics, Lehigh University, Bethlehem, PA, 18015, USA. {\tt\small \{vkp219, motee\}@lehigh.edu}.\endgraf
}
}

\maketitle

% Replace this two for submission
% \thispagestyle{plain}
% \pagestyle{plain}

\thispagestyle{empty}
\pagestyle{empty}

\begin{abstract}
Safe operation of connected vehicle platoons under stochastic disturbances and time-delayed dynamics requires accurate quantification of rare but dangerous events, such as inter-vehicle collisions.
We propose a rigorous framework for quantifying the risk of inter-vehicle collisions in connected vehicle platoons subject to time-delayed stochastic dynamics. We adopt the \emph{entropic value-at-risk} (EVaR) as a conservative metric to capture \emph{risk due to extreme events}, highlighting its advantages over conventional Value-at-Risk (VaR) and Conditional Value-at-Risk (CVaR). By expressing the inter-vehicle distance covariance in terms of the Laplacian eigenvalues of the communication network, we derive \emph{network-and time-delay-induced bounds} on both the minimum inherent risk and the worst-case risk. Specifically, the algebraic connectivity dictates the maximum EVaR, while the largest Laplacian eigenvalue determines the minimum risk inherently induced by the network structure. Numerical simulations illustrate how network topology and time delay shape collision risk, offering actionable insights for the safe design of vehicle platoons operating under stochastic disturbances.
\end{abstract}

%%%%%%%%%%%%%%%%%%%%%%%%%%%%%%%%%%%%%%%%%%%%%%%%%%%%%%%%%%%%%%%%%%%%%%%%%%%%%%%%%%%%%

\section{Introduction}

Connected vehicle platoons promise improved traffic throughput, fuel efficiency, and safety by coordinating vehicle motion through communication networks. However, practical deployments must operate under communication delays and stochastic disturbances arising from sensing noise, environmental effects, and actuation uncertainty. While existing platoon control strategies primarily focus on stability and performance guarantees, safe operation ultimately depends on understanding the probability of rare but catastrophic events such as inter-vehicle collisions. These extreme events are particularly challenging to analyze because their likelihood depends jointly on network topology, communication delays, and stochastic disturbances. Consequently, developing principled methods to quantify and conservatively bound collision risk in stochastic platoon dynamics remains a critical challenge for the safe design of networked vehicle systems.

% \section{Introduction}
% Connected vehicle platoons promise improved traffic throughput, fuel efficiency, and safety by coordinating vehicle motion through communication networks. However, practical deployments must operate under communication delays and stochastic disturbances arising from sensing noise, environmental effects, and actuation uncertainty. While existing platoon control strategies primarily focus on stability and performance guarantees, safe operation ultimately depends on understanding the probability of rare but catastrophic events such as inter-vehicle collisions. These extreme events are particularly challenging to analyze because their likelihood depends jointly on network topology, communication delays, and stochastic disturbances. Consequently, developing principled methods to conservatively quantify collision risk in stochastic platoon dynamics remains a critical challenge for the safe design of networked vehicle systems.

The presence of uncertainty in networked systems has motivated robustness and risk analysis in complex dynamical systems such as financial networks \cite{acemoglu2015systemic}, power networks \cite{Somarakis2018d}, integrated cyber-physical systems \cite{zhang2019robustness}, linear consensus networks \cite{Somarakis2017a}, and vehicle networks \cite{grunberg2017determining}. In the context of platoons, \cite{Somarakis2020b} uses Value-at-Risk (VaR) to quantify collision risk. This was extended to  Conditional Value-at-Risk (CVaR) \cite{liu2021risk} and, more recently, to distributionally robust formulations \cite{pandey2023quantification}. However, VaR \cite{artzner1999coherent} and CVaR \cite{rockafellar2002conditional} may \emph{underestimate tail risk} in safety-critical systems: VaR captures only a quantile, while CVaR averages the tail. This motivates the need for more conservative risk measures that rigorously account for extreme events.

To address this challenge, we adopt the \emph{entropic value-at-risk} (EVaR) \cite{ahmadi_javid2012evar}, which provides the tightest Chernoff-bound-based risk measure. EVaR conservatively quantifies the probability of extreme events by exponentially weighting the tail of the distribution, yielding an upper bound on both VaR and CVaR. This makes it particularly suitable for capturing the risk of inter-vehicle collisions in stochastic platoon dynamics under communication delays.
\vspace{0.25cm}
\begin{figure}[t]
\centering
\includegraphics[width=\linewidth]{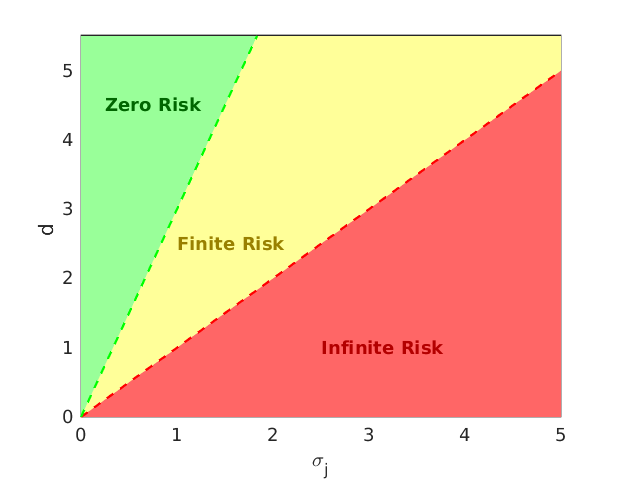}
\caption{Risk Geometry: Design Trade off Between Inter-vehicle Spacing and Network Connectivity through variance.}
\label{fig:d_vs_sigma_slope_diagram}
\end{figure}

\noindent\textbf{Contributions.} The main contributions of this paper are summarized as follows:
\begin{itemize}
    \item \textit{EVaR-Based Risk Quantification:} We introduce EVaR as a \emph{conservative and theoretically sound metric} for inter-vehicle collision risk in platoons subject to stochastic disturbances and time-delayed dynamics.
    \item \textit{Closed-Form Risk Formulation:} Using the \emph{moment-generating function and Chernoff bounds}, we derive an explicit formula for the EVaR-based risk of inter-vehicle collisions.
    \item \textit{Network- and Delay-Dependent Bounds:} We provide explicit bounds on both the \emph{worst-case} and \emph{minimum inherent collision risk}, showing that the \emph{algebraic connectivity} (second-smallest Laplacian eigenvalue) governs the worst-case EVaR, while the \emph{largest eigenvalue} dictates the minimum inherent risk. These results inform \emph{risk-resilient network design}.
    \item \textit{Simulation Studies:} Numerical simulations illustrate how \emph{network topology and communication delay} jointly influence collision risk, providing actionable insights for the safe design of connected vehicle platoons.
\end{itemize}

\vspace{0.25cm}
\noindent\textbf{Notation and Graph Preliminaries.}
Let $\{\bm e_1,\ldots,\bm e_n\}$ denote the standard Euclidean basis and 
$\bm 1_n$ the vector of ones. Define $\tilde{\bm e}_i=\bm e_{i+1}-\bm e_i$ for 
$i=1,\ldots,n-1$. A random vector $\bm y\in\mathbb R^q$ with mean $\bm\mu$ 
and covariance $\Sigma$ is denoted by $\bm y\sim\mathcal N(\bm\mu,\Sigma)$. 
The communication network is modeled by a connected undirected weighted graph 
$\mathcal G=(\mathcal V,\mathcal E,\omega)$. Its Laplacian matrix $L$ is symmetric 
positive semidefinite \cite{van2010graph} with eigenvalues
\[
0=\lambda_1<\lambda_2\le\cdots\le\lambda_n .
\]
Let $L=Q\Lambda Q^T$ denote its spectral decomposition with orthonormal 
eigenvectors $Q=[\bm q_1|\cdots|\bm q_n]$ and $\bm q_1=\frac{1}{\sqrt n}\bm 1_n$.

%%%%%%%%%%%%%%%%%%%%%%%%%%%%%%%%%%%%%%%%%%%%%%%%%%%%%%%%%%%%%%%%%%%%%%%%%%%%%%%%%%%%%

\section{Problem Formulation}\label{sec:problem_formulation}
\begin{figure}[t]
    \centering
	\includegraphics[width=\linewidth]{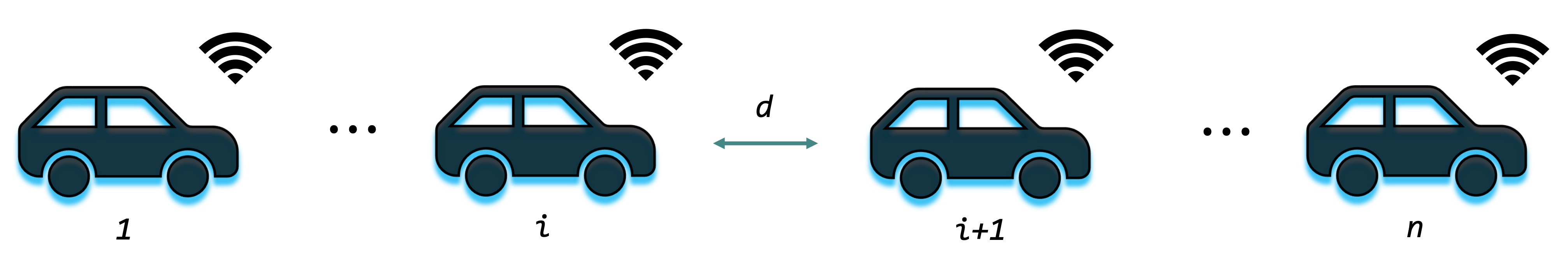}
	\caption{Schematic of a platoon where vehicles maintain a desired distance $d$ using delayed feedback over a communication network.}
	\label{fig:platoon_schematics}
\end{figure}
We consider a platoon of $n$ vehicles moving along a straight line (Fig.~\ref{fig:platoon_schematics}). 
Vehicles are indexed in descending order, where the $n$th vehicle acts as the leader. 
The state of vehicle $i$ is $[x_t^{(i)},v_t^{(i)}]^T$, where $x_t^{(i)}$ and $v_t^{(i)}$ denote position and velocity. 
The vehicle dynamics are modeled as
\begin{equation}
\begin{aligned}
d x_t^{(i)} &= v_t^{(i)} dt,\\
d v_t^{(i)} &= u_t^{(i)} dt + g\, d\xi_t^{(i)},
\end{aligned}
\label{eq:dyn_vehicle}
\end{equation}
where $u_t^{(i)}$ is the control input and $d\xi_t^{(i)}$ denotes standard Brownian motion representing environmental disturbances with diffusion coefficient $g>0$.

Vehicles aim to maintain a constant inter–vehicle spacing $d$ and achieve consensus in velocity. 
To this end, the feedback control law \cite{yu2010some} is
\begin{multline}
u_t^{(i)} =
\sum_{j=1}^n k_{ij}\big(v_{t-\tau}^{(j)}-v_{t-\tau}^{(i)}\big)
\\
+\beta\sum_{j=1}^n k_{ij}\big(x_{t-\tau}^{(j)}-x_{t-\tau}^{(i)}-(d_j-d_i)\big),
\label{eq:control_law}
\end{multline}
where $\beta>0$ weights the relative importance of velocity and position feedback, and $\tau>0$ denotes the communication delay.

Define the stacked state vectors
\[
\bm{x}_t=[x_t^{(1)},\ldots,x_t^{(n)}]^T,\quad
\bm{v}_t=[v_t^{(1)},\ldots,v_t^{(n)}]^T,
\]
and disturbance vector $\bm{\xi}_t=[\xi_t^{(1)},\ldots,\xi_t^{(n)}]^T$. 
Let $\bm{y}=[d,2d,\ldots,nd]^T$ denote the desired spacing from the leader.

Substituting \eqref{eq:control_law} into \eqref{eq:dyn_vehicle} yields the closed-loop platoon dynamics
\begin{equation}
\begin{aligned}
d\bm{x}_t &= \bm{v}_t dt,\\
d\bm{v}_t &= -L\bm{v}_{t-\tau}dt
-\beta L(\bm{x}_{t-\tau}-\bm{y})dt
+g\, d\bm{\xi}_t,
\end{aligned}
\label{eq:dyn_platoon}
\end{equation}
where $L$ is the Laplacian of the communication graph. 
Given deterministic initial conditions on $[-\tau,0]$, the stochastic delay differential equation \eqref{eq:dyn_platoon} admits a well-defined solution \cite{mohammed1984stochastic,Evans2013}.

\textbf{Problem Statement.}
We quantify the risk of inter-vehicle collision induced by noise and communication delays. 
Specifically, using the steady-state statistics of \eqref{eq:dyn_platoon}, we develop an entropic value-at-risk (EVaR) measure to characterize conservative bounds on collision risk as a function of the graph Laplacian, delay $\tau$, and disturbance statistics. 
All proofs are provided in the Appendix (extended version).

%%%%%%%%%%%%%%%%%%%%%%%%%%%%%%%%%%%%%%%%%%%%%%%%%%%%%%%%%%%%%%%%%%%%%%%%%%%%%%%%%%%%%
\section{Preliminary Results}\label{sec:prelims}

We briefly summarize the stability conditions and steady–state statistics of the platoon that are used in the subsequent risk analysis.

\subsection{Stability of the Deterministic Platoon}

The platoon is said to converge if the velocity of all vehicles and the inter–vehicle distances converge to constant values $v$ and $d$, respectively, i.e.,
\[
\lim_{t\to\infty}|v_t^{(i)}-v_t^{(j)}|=0, \qquad
\lim_{t\to\infty}|x_t^{(i)}-x_t^{(j)}-(i-j)d|=0 .
\]

Using the results in \cite{yu2010some,bellman1963_dde}, the deterministic platoon converges if and only if
\begin{equation}
(\lambda_i\tau,\beta\tau)\in S, \qquad i=2,\ldots,n ,
\label{eq:platoon_stability}
\end{equation}
where $\lambda_i$ are the Laplacian eigenvalues and
\begin{equation*}
    \scalebox{0.91}{$S=\Big\{(s_1,s_2)\in\mathbb{R}^2 \mid s_1\in(0,\tfrac{\pi}{2}), \;
s_2\in\big(0,\tfrac{a}{\tan a}\big), \;
a\sin a=s_1 \Big\}$}.
\end{equation*}
% \[
% S=\Big\{(s_1,s_2)\in\mathbb{R}^2 \mid s_1\in(0,\tfrac{\pi}{2}), \;
% s_2\in\big(0,\tfrac{a}{\tan a}\big), \;
% a\sin a=s_1 \Big\}.
% \]

Throughout the paper we assume the stability condition \eqref{eq:platoon_stability} holds.

\subsection{Steady-State Statistics of Inter-Vehicle Distances}

Let $L=Q\Lambda Q^T$ denote the spectral decomposition of the graph Laplacian and consider the coordinate transformation
\begin{equation}
\bm z_t=Q^T(\bm x_t-\bm y), \qquad 
\bm v_t=Q^T\bm v_t .
\label{eq:trans_pushforward}
\end{equation}

In these coordinates the platoon dynamics decouple as
\begin{equation}
\begin{aligned}
d\bm z_t &= \bm v_t dt,\\
d\bm v_t &= -\Lambda \bm v_{t-\tau} dt
           -\beta\Lambda(\bm z_{t-\tau}-\bm v)dt
           + gQ^T d\bm\xi_t .
\end{aligned}
\label{eq:dyn_platoon_decoupled}
\end{equation}

Under the stability condition \eqref{eq:platoon_stability}, the solution admits the representation
\begin{equation}
\begin{bmatrix}z_t^{(i)}\\ v_t^{(i)}\end{bmatrix}
=\Xi(\cdot)+
g\int_0^t \Phi_i(t-s)B_i\,d\bm\xi_s ,
\label{eq:dyn_platoon_decoupled_solution}
\end{equation}
where $B_i=[0_{1\times n},q_i]^T$ and $\Phi_i(t)$ is the principal solution of the deterministic system \cite{Somarakis2020b}. For $i=2,\ldots,n$, both $\Phi_i$ and $\Xi$ decay exponentially.

Consequently, the steady-state transformed state satisfies
\[
\Bar{\bm z}\sim \mathcal N(0,\Sigma_z),
\]
with diagonal covariance
\[
\Sigma_z=\text{diag}\{\sigma_{z_1}^2,\ldots,\sigma_{z_n}^2\},\qquad
\sigma_{z_i}^2=\frac{g^2\tau^3}{2\pi}f(\lambda_i\tau,\beta\tau),
\]
where
\begin{equation}
f(s_1,s_2)=
\int_{\mathbb R}
\frac{dr}{(s_1s_2-r^2\cos r)^2+r^2(s_1-r\sin r)^2}.
\label{eq:f(S_1,s_2)}
\end{equation}

Transforming back to the original coordinates,
\begin{equation}
\bm x_t=Q\bm z_t+\bm y .
\label{eq:trans_pullback}
\end{equation}

Define the steady-state distance vector $\Bar{\bm d}\in\mathbb R^{n-1}$ as 
\begin{equation}
\Bar{\bm d}=D^TQ\Bar{\bm z}+d\bm 1_{n-1},
\label{eq: steady_state_dist}
\end{equation}
where $D=[\tilde{\bm e}_1|\cdots|\tilde{\bm e}_{n-1}]$. Then $\Bar{\bm d}\sim\mathcal N(d\bm 1_{n-1},\Sigma)$. 
The marginal variances denoted
$\sigma_i^2=\Sigma(i,i)$ is given by
\begin{equation} 
\sigma_i^2
=
g^2\frac{\tau^3}{2\pi}
\sum_{k=2}^n
(\tilde{\bm e}_i^T q_k)^2
f(\lambda_k\tau,\beta\tau),
\label{eq:steady_state_var}
\end{equation}
for $i=1,\ldots,n-1$.

% where $D=[\tilde{\bm e}_1|\cdots|\tilde{\bm e}_{n-1}]$. Then 
% $\Bar{\bm d}\sim\mathcal N(d\bm 1_{n-1},\Sigma)$ \cite{liu2021risk}. 
% Since only the marginal variances are required in the sequel, we denote 
% $\sigma_i^2=\mathrm{Var}(\Bar{\bm d}_i)$, which is given by
% \begin{equation}
% \sigma_i^2 =
% g^2\frac{\tau^3}{2\pi}
% \sum_{k=1}^{n}
% (\tilde{\bm e}_i^T q_k)^2
% f(\lambda_k\tau,\beta\tau),
% \label{eq:steady_state_var}
% \end{equation}
% for $i=1,\ldots,n-1$.

% Then
% \[
% \Bar{\bm d}\sim\mathcal N(d\bm 1_{n-1},\Sigma),
% \]
% where
% \begin{equation}
% \Sigma=D^TQ\,\mathbb E[\Bar{\bm z}\Bar{\bm z}^T]Q^TD,
% \label{eq:sigma_transform}
% \end{equation}
% whose elements are
% \begin{equation}
% \sigma_{ij}
% =
% g^2\frac{\tau^3}{2\pi}
% \sum_{k=1}^n
% (\tilde{\bm e}_i^T q_k)(\tilde{\bm e}_j^T q_k)
% f(\lambda_k\tau,\beta\tau).
% \label{eq:steady_state_cov}
% \end{equation}

% For simplicity we denote $\sigma_{ii}$ by $\sigma_i^2$.

\section{Entropic Value-at-Risk Measure}

In this section, we construct a mathematical framework for characterizing extreme events and introduce risk measures used to quantify the likelihood of such events. We begin by defining the notion of an extreme event.

\subsection{Extreme Event}

A \emph{systemic event} refers to a failure that may lead to a global malfunction of the network. In the context of vehicle platoons, this corresponds to an inter-vehicle collision. For a vehicle pair $j$, such a failure occurs when the inter-vehicle distance satisfies $\bar{d}_j \to 0$. In the space of random variables, this event can be represented by 
\[
\mathcal{W}^* = (-\infty, \gamma),
\]
where $\gamma \to 0^{+}$. 

Risk analysis is meaningful only before such an event occurs; therefore, the objective is to quantify how close the system operates to this systemic failure so that appropriate design choices can prevent it. To this end, we introduce a family of nested level sets $\{\mathcal{W}_{\delta} \mid \delta \in [0,\infty]\}$ that approximate $\mathcal{W}^*$ and satisfy the following properties for any sequence $\{\delta_n\}_{n=1}^{\infty}$ with $\lim_{n\to\infty} \delta_n = \infty$:
\begin{itemize}
    \item $\mathcal{W}_{\delta_2} \subset \mathcal{W}_{\delta_1}$ whenever $\delta_1 < \delta_2$,
    \item $\lim_{n\to\infty} \mathcal{W}_{\delta_n} = \bigcap_{n=1}^{\infty} \mathcal{W}_{\delta_n} = \mathcal{W}^*$.
\end{itemize}

We model these extreme-event sets as
\begin{equation}
\mathcal{W}_{\delta} = (-\infty, \alpha(\delta)),
\end{equation}
where
\[
\alpha(\delta) = \frac{d}{\delta + c},
\]
and $c$ is chosen so that $\delta = 0$ corresponds to $\bar{d}_j = d/c$. Without loss of generality, we assume $c \geq 1$. 

The parameter $\delta$ therefore quantifies the proximity of the inter-vehicle distance $\bar{d}_j$ to the systemic event $\mathcal{W}^*$. Larger values of $\delta$ correspond to increasingly severe realizations approaching collision. Having defined the extreme-event structure, we now introduce the risk measures used to quantify the probability of such events.

\subsection{Risk Measures}

To quantify the tail risk associated with extreme events, we consider several widely used risk measures. While this work primarily focuses on the \emph{Entropic Value-at-Risk} (EVaR), we also introduce \emph{Value-at-Risk} (VaR) and \emph{Conditional Value-at-Risk} (CVaR) as benchmarks for comparison in both the theoretical analysis and simulations. These measures provide progressively stronger descriptions of extreme outcomes: VaR identifies a threshold exceeded with small probability, CVaR captures the expected magnitude beyond that threshold, and EVaR provides the tightest exponential upper bound on tail probabilities.

Unlike classical formulations that quantify risk proportional to the deviation of random variable, our objective is to measure the risk of \emph{collision} arising from extreme events. Consequently, the risk measures are adapted so that risk increases as the inter-vehicle distance decreases. In particular, the risk measures are constructed to be monotonically decreasing functions of the distance $\bar{d}_j$: as $\bar{d}_j \to 0$, the risk diverges. Moreover, rather than measuring risk in the physical units of distance, we express risk as a \emph{dimensionless quantity}, allowing comparisons across different network configurations and design parameters.

\begin{definition}\label{def:var_cvar}
Let \(d_j\) denote the inter-vehicle distance of the \(j\)-th vehicle pair, and let \(\varepsilon \in (0,1)\) be a confidence parameter. The \emph{Value-at-Risk} (VaR) at level \(\varepsilon\) is defined as
\begin{equation}\label{eqn:var}
V_{\varepsilon}(\bar{d}_j) := \inf \left\{ \delta \ge 0 \mid \mathbb{P}\big(\bar{d}_j \in \mathcal{W}_{\delta}\big) \le \varepsilon \right\},
\end{equation}
which represents the smallest threshold such that the probability of the extreme event does not exceed \(\varepsilon\). 

The \emph{Conditional Value-at-Risk} (CVaR) at the same level is defined as
\begin{equation}\label{eqn:cvar}
C_{\varepsilon}(\bar{d}_j) := \inf \left\{ \delta \ge 0 \mid 
\alpha(\delta) - \mathbb{E}\big[\bar{d}_j \mid \bar{d}_j \ge \alpha(V_\varepsilon)\big] \le 0 \right\},
\end{equation}
which captures the expected magnitude of extreme events beyond the VaR threshold.
\end{definition}

While VaR and CVaR provide useful summaries of tail behavior, they do not fully characterize the distribution of extreme losses. The \emph{Entropic Value-at-Risk} (EVaR) addresses this limitation by exploiting the \emph{moment generating function} (MGF), thereby providing the tightest exponential bound on tail probabilities.

Before defining EVaR, we recall the definition of the moment generating function.

\begin{definition}[Moment Generating Function {\cite{Kobayashi2011}}]\label{def:mgf}
Let \(X\) be a real-valued random variable. The \emph{moment generating function} (MGF) of \(X\), denoted \(M_X(s)\), is defined as
\begin{equation}\label{eqn:mgf}
M_X(s) := \mathbb{E}\big[e^{sX}\big],
\end{equation}
for all \(s\) such that the expectation exists and is finite. The MGF characterizes all moments of \(X\) and plays a central role in describing tail behavior.
\end{definition}

To capture proximity of inter-vehicle distance $(\bar{d}_j)$ to the systemic event of inter-vehicle collision $(\bar{d}_j \to 0)$, we employ the \emph{Entropic Value-at-Risk} (EVaR) \cite{ahmadi_javid2012evar}, which provides a conservative, coherent, and tail-sensitive measure of risk.

\begin{lemma}\label{lem:evar_def}
Let \(\bar{d}_j\) be a random variable whose moment generating function \(M_{\bar{d}_j}(s)\) exists for all \(s > 0\), and let \(\varepsilon \in (0,1)\). The \emph{Entropic Value-at-Risk} of \(\bar{d}_j\) at level \(\varepsilon\) is defined as
\begin{equation}\label{eqn:evar}
E_{\varepsilon}(\bar{d}_j) := 
\inf 
\left\{
\delta \geq 0  \;\middle|\;
\inf_{s>0}
\left[
\alpha(\delta) \, s + \ln M_{\bar{d}_j}(-s)
\right]
\le \ln \varepsilon
\right\}.
\end{equation}
\end{lemma}

In this formulation, the auxiliary parameter \(s\) exponentially tilts the distribution toward realizations with small inter-vehicle distances. The inner optimization selects the tightest exponential bound over all such tilts, yielding the smallest exponential upper bound on the probability of extreme proximity events. Hence, EVaR characterizes the left-tail behavior of \(\bar{d}_j\), which corresponds to collision risk.

% \begin{remark}[Entropic interpretation]
% The definition in Lemma~\ref{lem:evar_def} can be interpreted through the dual representation of EVaR \footnote{This follows from Donsker-Varadhan Variational Formula \cite{Dupuis1997donsker_varadhan}. $\ln \mathbb{E}_\mathbb{P}(e^X) = \sup_{Q<<P} \{\mathbb{E}_\mathbb{Q}(X) - D_{KL}(\mathbb{Q} \mid \mid \mathbb{P})\}$ . Due to page limit, the analysis has been omitted.}
% \[
% \mathrm{E}_\varepsilon(\bar{d}_j)
% =
% \sup_{Q\ll P}
% \left\{
% \mathbb E_Q[-\bar{d}_j]
% \mid
% D_{\mathrm{KL}}(Q\|P)\le -\ln(\varepsilon)
% \right\},
% \]
% where $D_{\mathrm{KL}}$ denotes the Kullback–Leibler divergence. 
% In the present setting this implies that EVaR characterizes the worst-case expected inter-vehicle proximity under probability distributions that remain close to the nominal one in the sense of relative entropy.
% \end{remark}

\begin{proposition}[Entropic interpretation] \label{prop:evar_dual}
The definition in Lemma~\ref{lem:evar_def} admits a dual representation via the Donsker–Varadhan variational formula \cite{Dupuis1997donsker_varadhan}\footnote{For a random variable $X$, $\ln \mathbb{E}_\mathbb{P}[e^X] = \sup_{Q \ll P} \{\mathbb{E}_Q[X] - D_{\mathrm{KL}}(Q\|P)\}$. Analysis omitted due to space constraints.}:
% \[
% \mathrm{E}_\varepsilon(\bar{d}_j)
% = \inf \{\delta \geq 0 \mid \alpha(\delta) + 
% \sup_{Q\ll P}
% \left\{
% \mathbb E_Q[-\bar{d}_j]
% \;\middle|\;
% D_{\mathrm{KL}}(Q\|P)\le -\ln(\varepsilon)
% \right\}\},
% \]
\[
\mathrm{E}_\varepsilon(\bar d_j)
=
\inf
\left\{
\delta \ge 0
\;\middle|\;
\alpha(\delta) + \Psi_\varepsilon(\bar d_j) \le 0
\right\},
\]
where 
\[
\Psi_\varepsilon(\bar d_j)
:=
\sup_{Q\ll P}
\left\{
\mathbb E_Q[-\bar d_j]
:\,
D_{\mathrm{KL}}(Q\|P)\le -\ln\varepsilon
\right\}, 
\] and
$D_{\mathrm{KL}}$ denotes the Kullback–Leibler divergence. 
\end{proposition}
This dual representation reveals that EVaR has an entropy-based
distributionally robust interpretation, where the risk corresponds
to the worst-case expected inter-vehicle proximity over probability
distributions lying within a relative-entropy (KL-divergence) ball
around the nominal distribution.
% This representation shows that EVaR admits an entropy-based
% distributionally robust interpretation: the risk corresponds to the
% worst-case expected inter-vehicle proximity under probability
% distributions that remain within a relative-entropy
% (Kullback--Leibler divergence) neighborhood of the nominal
% distribution.
% Thus, EVaR admits a distributionally robust interpretation, where the risk
% measure corresponds to the worst-case expected proximity under probability
% distributions lying in a KL-divergence ball around the nominal distribution.

% In this context, EVaR quantifies the worst-case expected inter-vehicle proximity under alternative distributions that remain close to the nominal distribution in the sense of relative entropy, and it naturally increases as $\bar{d}_j$ decreases.
% \end{proposition}

\begin{remark}\label{rem:var_cvar_evar_order}
For any \(\varepsilon \in (0,1)\), the risk measures satisfy the ordering
\begin{equation}\label{eqn:var_cvar_evar_order}
V_\varepsilon(\bar{d}_j) \le C_\varepsilon(\bar{d}_j) \le E_\varepsilon(\bar{d}_j).
\end{equation}
\end{remark}

This ordering highlights the increasing tail sensitivity of the three measures. While VaR captures a quantile threshold and CVaR measures the expected loss beyond that threshold, EVaR accounts for the \emph{entire tail distribution} through its moment-generating-function representation, resulting in the most conservative coherent risk measure among the three.

%%%%%%%%%%%%%%%%%%%%%%%%%%%%%%%%%%%%%%%%%%%%%%%%%%%%%%%%%%%%%%%%%%%%%%%%%%%%%%%%%%%%%

\section{Risk of Collision Due to Extreme Events}   \label{sec:risk}

In this section, we quantify the tail risk associated with large fluctuations in inter-vehicle distance. We first characterize the moment generating function (MGF) of the inter-vehicle distance, which will then be used to compute the \emph{Entropic Value-at-Risk} (EVaR).

% \subsection{Moment Generating Function of Inter-Vehicle Distance}
\subsection{Entropic Value-at-Risk of Collision}

\begin{lemma}   \label{lem:mgf_d_bar_j}
Let $\bar{d}_j \sim \mathcal{N}(d, \sigma_j^2)$. Then the moment generating function of $\bar{d}_j$ is
\begin{equation} \label{eqn:mgf_mod_y}
    M_{\bar{d}_j}(s) = e^{sd + \frac{\sigma_j^2 s^2}{2}} .
\end{equation}
\end{lemma}

Lemma \ref{lem:mgf_d_bar_j} provides a closed-form expression for the MGF of inter-vehicle distance, which is essential for computing EVaR using Lemma \ref{lem:evar_def}.

% \subsection{Entropic Value-at-Risk of Collision}

\begin{theorem}\label{thm:evar_inter_vehicle}
Let $\bar{d}_j$ denote the inter-vehicle distance between vehicles $j$ and $j+1$. Then the EVaR of collision due to extreme events is
\begin{equation} \label{eqn:evar_inter_vehicle_collision}
    E_{\varepsilon}(\bar{d}_j) = 
    \begin{cases}        
    0 , & \kappa_\varepsilon \geq \frac{c}{c -1} ,\\[1ex]
    \frac{1}{1 - \frac{1}{\kappa_\varepsilon}} - c , & 1 \leq \kappa_\varepsilon < \frac{c}{c - 1} ,\\[0.5ex]
    \infty , & \kappa_\varepsilon \leq 1,
    \end{cases}
\end{equation}
where 
\[
d_\varepsilon = \frac{d}{\sqrt{-\ln \varepsilon}}, \quad 
\kappa_\varepsilon = \frac{d_\varepsilon}{\sqrt{2} \, \sigma_j}.
\]
\end{theorem}

Theorem \ref{thm:evar_inter_vehicle} expresses EVaR as a dimensionless number $\kappa_\varepsilon$, capturing the tradeoff between mean inter-vehicle distance $d$ and its standard deviation $\sigma_j$.  
Case 1: Two distributions with the same mean but different variances show that risk depends on standard deviation, which is a function of network topology through \eqref{eq:steady_state_var}.  
Case 2: Distributions with different means and the same variance show that risk also depends on mean distance, which is a platoon design parameter.  
Thus, EVaR captures the interplay between network design (topology) and platoon design (steady-state distances). 

This leads to a fundamental tradeoff: for fixed design parameters $c$ and $\varepsilon$, risk is determined solely by the slope $\kappa_\varepsilon$.  
% - $\kappa_\varepsilon > \frac{c}{c-1}$: zero risk  
% - $1 \le \kappa_\varepsilon < \frac{c}{c-1}$: nonzero risk  
% - $\kappa_\varepsilon < 1$: infinite risk  
A plot of $d_\varepsilon$ versus $\sqrt{2}\sigma_j$ visualizes these three regions (Fig.~\ref{fig:d_vs_sigma_slope_diagram}), showing a cutoff slope that depends on both mean distance and variance. This provides a simple design guideline: adjusting network topology or platoon spacing can directly control risk.

\noindent

\subsection{Network-Induced Bounds on EVaR}

Since time delays are inherent to the system dynamics and cannot be altered, the two design parameters that affect collision risk are the \emph{network topology} and the \emph{platoon distance}. To this end, we derive network-induced bounds on the entropic value-at-risk, which quantify the \emph{maximum} and \emph{minimum} risk across all inter-vehicle distances in the platoon.

% For clarity, we introduce the following notation:  

\begin{theorem}[EVaR Bounds via Laplacian Eigenvalues]\label{thm:evar_laplacian_bounds}
Let $\bar{d}_j$ denote the inter-vehicle distance between the $j$-th and $(j+1)$-th vehicles in a platoon. Restricting to the middle branch of EVaR without loss of generality, assume
\[
\left( 1- \frac{1}{c}\right)^2 d_\varepsilon^2 \leq f(\lambda_n \tau, \beta \tau) \leq f(\lambda_2 \tau, \beta \tau) \leq d_\varepsilon^2.
\]

Then, the entropic value-at-risk satisfies
\begin{equation}
E_{\varepsilon}^{n} \leq \inf_{j \in \{1, \dots, n-1\}} E_{\varepsilon}(\bar{d}_j) \leq \sup_{j \in \{1, \dots, n-1\}} E_{\varepsilon}(\bar{d}_j) \leq E_{\varepsilon}^{2},
\end{equation}
where the \emph{minimum} and \emph{maximum} EVaR bounds are
\[
E_\varepsilon^\star = \frac{1}{1 - 1/\kappa_\varepsilon^\star}, \qquad \star \in \{2,n\},
\]
with
\[
\kappa_\varepsilon^\star = \frac{d_\varepsilon}{2 \sqrt{c_1 f(\lambda_\star\tau,\beta\tau)}}, \qquad
c_1 = g^2 \frac{\tau^3}{2 \pi}.
\]

These bounds provide a direct spectral interpretation of EVaR in platoons: the \emph{minimum risk} arises from the largest Laplacian eigenvalue ($\lambda_n$), while the \emph{worst-case risk} is determined by the algebraic connectivity ($\lambda_2$).  
\end{theorem}

These network-induced bounds are crucial for design: they allow us to minimize the \emph{worst-case risk} across all inter-vehicle distances while also understanding the \emph{inherent minimum risk} imposed by the network structure.

\begin{figure}[t]
\centering

\begin{subfigure}{0.32\linewidth}
\centering
\includegraphics[width=\linewidth]{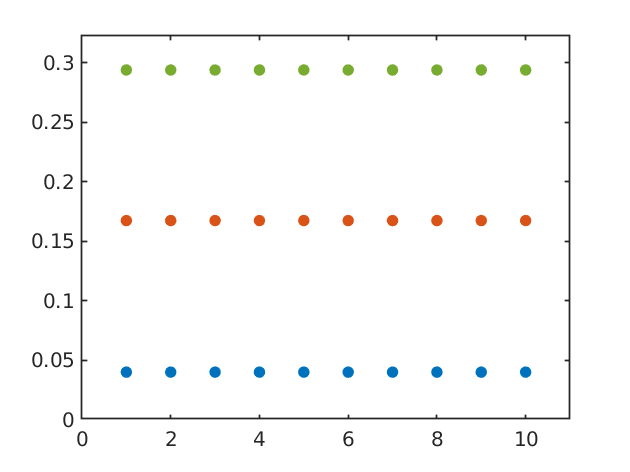}
\caption{}
\end{subfigure}
\hfill
\begin{subfigure}{0.32\linewidth}
\centering
\includegraphics[width=\linewidth]{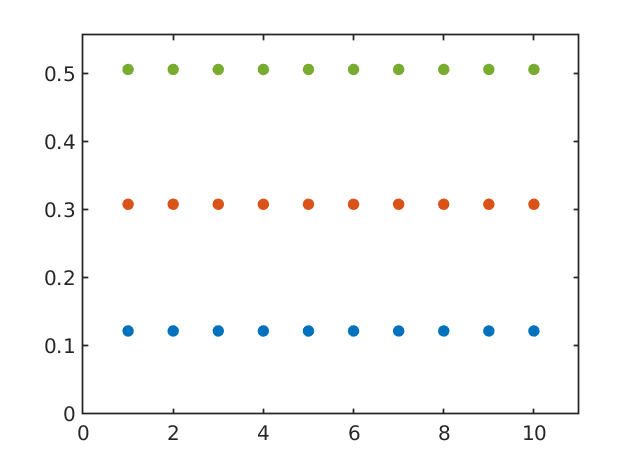}
\caption{}
\end{subfigure}
\hfill
\begin{subfigure}{0.32\linewidth}
\centering
\includegraphics[width=\linewidth]{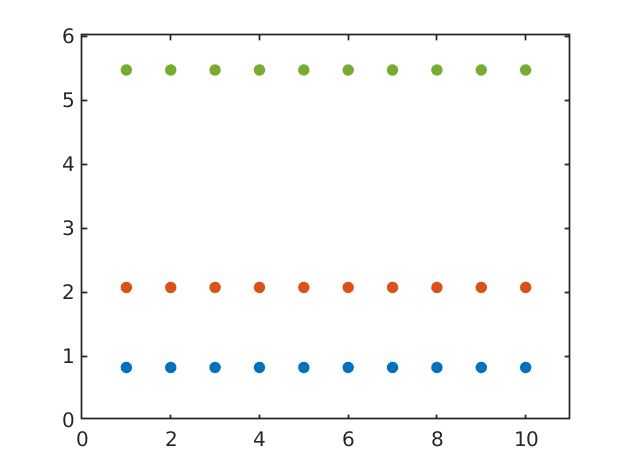}
\caption{}
\end{subfigure}
\vspace{0.3cm}
\begin{subfigure}{0.32\linewidth}
\centering
\includegraphics[width=\linewidth]{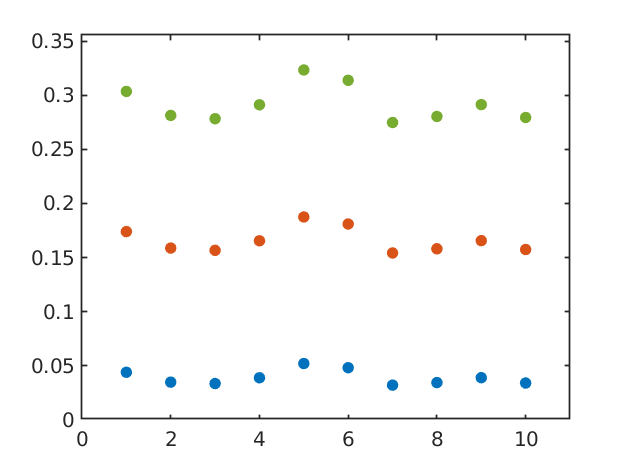}
\caption{}
\end{subfigure}
\hfill
\begin{subfigure}{0.32\linewidth}
\centering
\includegraphics[width=\linewidth]{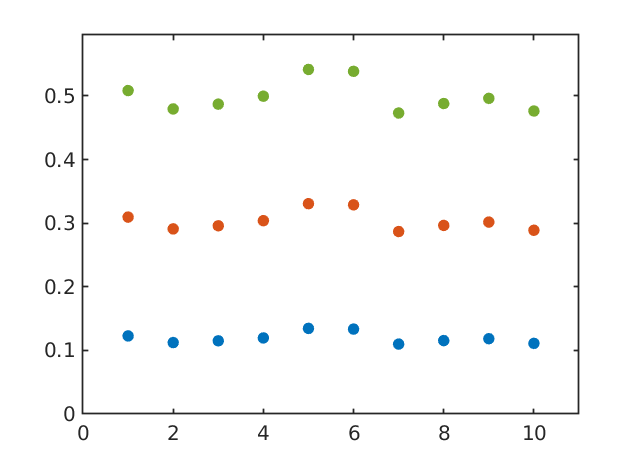}
\caption{}
\end{subfigure}
\hfill
\begin{subfigure}[b]{0.32\linewidth}
\centering
\includegraphics[width=\linewidth]{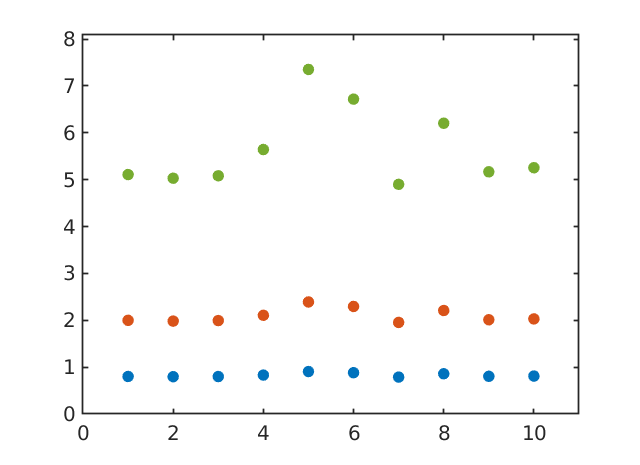}
\caption{}
\end{subfigure}

\begin{tabular}{ccc}
\makebox[0.28\linewidth]{Complete graph} &
\makebox[0.28\linewidth]{8-cycle graph} &
\makebox[0.28\linewidth]{4-cycle graph} \\
\end{tabular}

\caption{Comparison of VaR ({\color[HTML]{0072BD}$\bullet$}), CVaR ({\color[HTML]{D95319}$\bullet$}), and EVaR ({\color[HTML]{77AC30}$\bullet$}) risks across different graph topologies. $y-$ axis corresponds to risk values, while $x-$axis corresponds to intervehicle indices.
Top row: uniform edge weights $w=1$. Bottom row: random edge weights $w\in[0.8,1.2]$.}
\label{fig:risk_comparison_graph_topologies}
\end{figure}

% 1. Comapre Var , CVAr, EVar uniform weight.
% 2. Comapre them random weight. 
% 3. Compare epsilon = 0.05, 0.10,0.15
% 4. Try to get network bounds.

\section{Case Studies}\label{sec:case_studies}

We present several case studies for platoons whose dynamics are governed by \eqref{eq:dyn_platoon} under different communication graph topologies. Unless otherwise stated, the simulation parameters are chosen as
$n = 11$, $c = 1.21$, $d = 1.01$, $\tau = 0.01$, $\beta = 1/3$, and $g = 1$.

\subsection{Comparison of VaR, CVaR and EVaR}

We first compare the risk measures VaR, CVaR, and EVaR for different graph topologies. These quantities are computed using \eqref{eqn:var}, \eqref{eqn:cvar} (Definition \ref{def:var_cvar}), and \eqref{eqn:evar_inter_vehicle_collision} (Theorem \ref{thm:evar_inter_vehicle}), respectively. Equivalent formulations for VaR and CVaR can also be found in \cite{Somarakis2020b} and \cite{liu2025risk}. For this analysis, we consider $\varepsilon = 0.1$.

Two sets of experiments are conducted. In the first case, we consider graphs with uniform edge weights equal to $1$. In the second case, the edge weights are randomly selected in the interval $[0.8,1.2]$. For both cases, we analyze complete graphs and $p$-cycle graphs with varying connectivity levels. The results are shown in Fig.~\ref{fig:risk_comparison_graph_topologies}.

As observed from the figure, all three risk measures decrease as the connectivity of the communication graph increases. Among them, EVaR consistently yields the largest value, acting as a conservative upper bound on the risk as stated in Remark \ref{rem:var_cvar_evar_order}. This property highlights the importance of EVaR in safety-critical applications such as inter-vehicle collision avoidance in vehicle platoons.

We do not include the path graph in this comparison because all three risk measures diverge for this topology. A similar behavior is observed for the $2$-cycle graph. This phenomenon is explained next using the spectral bounds developed in Theorem \ref{thm:evar_laplacian_bounds}.

\subsection{Risk Bounds}

Beyond network design considerations, Theorem \ref{thm:evar_laplacian_bounds} provides insight into why certain graph topologies are inherently risky from a safety perspective.

\begin{figure}[t]
\centering

\begin{subfigure}{0.48\linewidth}
\centering
\includegraphics[width=\linewidth]{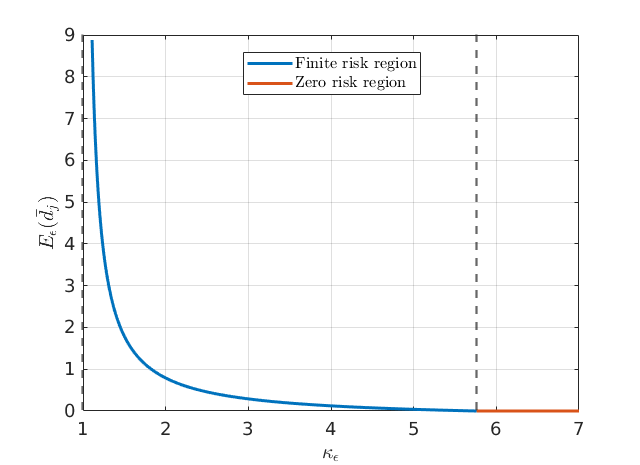}
\caption{EVaR vs $\kappa_\varepsilon.$}
\end{subfigure}
\hfill
\begin{subfigure}{0.48\linewidth}
\centering
\includegraphics[width=\linewidth]{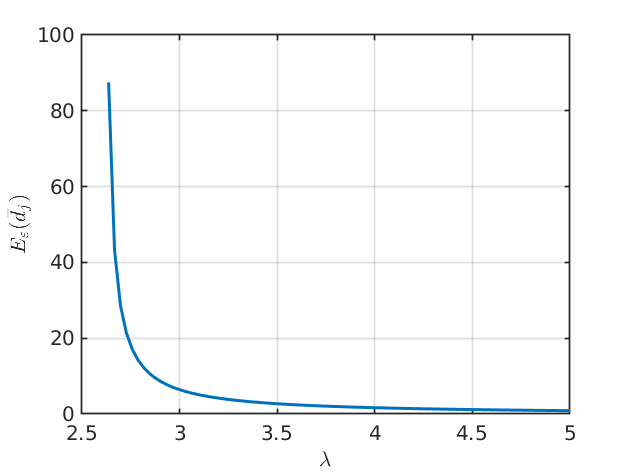}
\caption{EVar vs $\lambda$}
\end{subfigure}
\hfill

\caption{Network Induced Bounds on EVaR}
\label{fig:network_induced_bounds}
\end{figure}

To illustrate this, consider the unweighted path graph, whose maximum Laplacian eigenvalue is bounded by $2$. Using the simulation parameters specified earlier, a direct calculation shows that the resulting value of $\kappa_\varepsilon$ is less than $1$. According to Theorem \ref{thm:evar_inter_vehicle}, this implies that the EVaR becomes infinite, indicating an unacceptable collision risk. The EVaR profile as a function of $\kappa_\varepsilon$ is shown in Fig.~\ref{fig:network_induced_bounds}(a).

A similar conclusion holds for the $2$-cycle graph. This behavior can also be interpreted from Fig.~\ref{fig:network_induced_bounds}(b), where the risk approaches a vertical asymptote as the Laplacian eigenvalue approaches approximately $2.5$. This demonstrates that even the minimum achievable risk for these graph topologies remains unbounded.

\subsection{EVaR vs $\varepsilon$}
\begin{figure}[t]
\centering

\begin{subfigure}{0.48\linewidth}
\centering
\includegraphics[width=\linewidth]{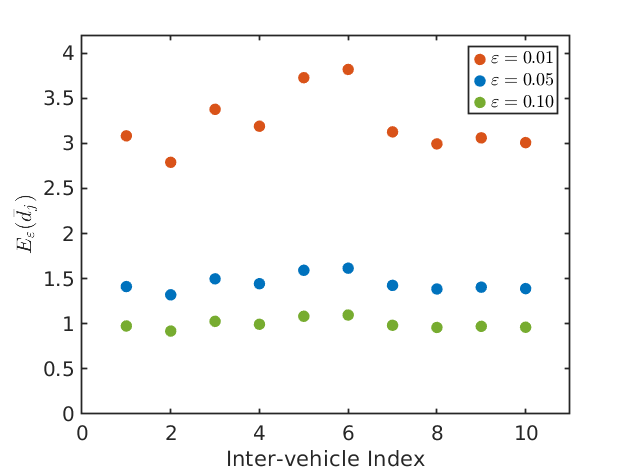}
\caption{EVaR profile for 6-cycle graph for different values of $\varepsilon$}
\end{subfigure}
\hfill
\begin{subfigure}{0.48\linewidth}
\centering
\includegraphics[width=\linewidth]{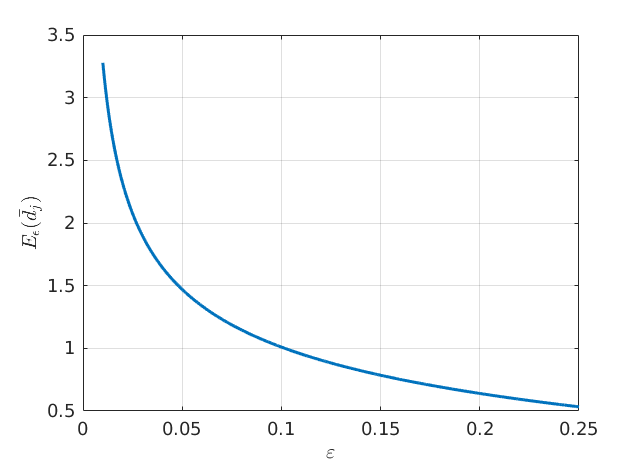}
\caption{EVar vs $\varepsilon$ for unweighted 6 cycle graph}
\end{subfigure}
\hfill

\caption{EVaR vs $\varepsilon$}
\label{fig:evar_vs_epsilon}
\end{figure}

Finally, we investigate how the EVaR varies as the parameter $\varepsilon$ changes. To study this effect, we consider a randomly weighted $6$-cycle graph with edge weights uniformly sampled from the interval $[0.8,1.2]$. The resulting risk profile is shown in Fig.~\ref{fig:evar_vs_epsilon}(a). As $\varepsilon$ decreases, the EVaR associated with each inter-vehicle distance increases significantly.

To further illustrate this trend, Fig.~\ref{fig:evar_vs_epsilon}(b) shows a continuous EVaR profile as a function of $\varepsilon$ for an unweighted $6$-cycle graph. The figure clearly demonstrates that EVaR increases as $\varepsilon$ decreases, indicating an inverse relationship between the risk measure and the confidence parameter.

This behavior also highlights the distributionally robust interpretation of EVaR, as characterized by the dual representation in Proposition~\ref{prop:evar_dual}. In particular, decreasing $\varepsilon$ enlarges the radius of the Kullback--Leibler (KL) divergence ball. Consequently, the expectation is evaluated over a larger set of probability distributions, leading to a more conservative (larger) risk value.

\section{Conclusion} \label{sec:conclusion}

In this paper, we developed a rigorous framework for quantifying the risk of inter-vehicle collisions in connected vehicle platoons subject to stochastic disturbances and communication delays. We proposed the use of the entropic value-at-risk (EVaR) as a principled metric for capturing collision risk due to extreme events. Unlike conventional risk measures such as VaR and CVaR, EVaR provides a tighter and more conservative characterization of tail events, making it particularly suitable for safety-critical systems such as vehicle platoons.
By exploiting the spectral properties of the communication network, we derived explicit bounds on the EVaR of inter-vehicle collisions induced by both network topology and time delays. 
% In particular, we showed that the algebraic connectivity of the network governs the worst-case collision risk, while the largest Laplacian eigenvalue determines the minimum inherent risk imposed by the network structure. This spectral characterization provides a clear and interpretable link between network connectivity and safety in platoons.
 Numerical simulations illustrated how the interplay between network connectivity and delay shapes Evar across entire network.
Future work will investigate the design of network topologies and control strategies that explicitly minimize EVaR-based collision risk, as well as extensions to cascade of inter-vehicle collisions in the platoon.
% \subsection{Worst case Risk}

% \begin{figure}[t]
% \centering

% \begin{subfigure}{0.48\linewidth}
% \centering
% \includegraphics[width=\linewidth]{Figs/evar_cvar_var_complete_w_rand_0.8_1.2.png}
% \caption{Complete graph}
% \end{subfigure}
% \hfill
% \begin{subfigure}{0.48\linewidth}
% \centering
% \includegraphics[width=\linewidth]{Figs/evar_cvar_var_8_cycle_w_rand_0.8_1.2.png}
% \caption{8-cycle graph}
% \end{subfigure}

% \vspace{0.3cm}

% \begin{subfigure}{0.48\linewidth}
% \centering
% \includegraphics[width=\linewidth]{Figs/evar_cvar_var_6_cycle_w_rand_0.8_1.2.png}
% \caption{6-cycle graph}
% \end{subfigure}
% \hfill
% \begin{subfigure}{0.48\linewidth}
% \centering
% \includegraphics[width=\linewidth]{Figs/evar_cvar_var_4_cycle_w_rand_0.8_1.2.png}
% \caption{4-cycle graph}
% \end{subfigure}

% \caption{Comparison of VaR, CVaR, and EVaR risks across different graph topologies for random edge weights between to 0.8 - 1.2.}
% \label{fig:risk_comparison_uniform_weights_w_rand_0.8_1.2}
% \end{figure}

% \begin{figure}[t]
% \centering
% \includegraphics[width=0.75\linewidth]{Figs/E_max_vs_lambda_2_v1.png}
% \caption{Worst case Risk vs Algebraic connectivity}

% \label{fig:worst_risk_vs_algebraic_connectivity}
% \end{figure}

\printbibliography

 \newpage

\appendix

\subsubsection{Proof of Lemma~\ref{lem:evar_def}}

The proof follows the approach in \cite{ahmadi_javid2012evar}, adapted to characterize collision risk arising from small inter-vehicle distances.

Using the Chernoff bound for the left tail \cite{Boucheron2013ConcentrationIA}, for any $s > 0$ we obtain
\begin{align*}
\mathbb{P}\!\left(\bar d_j \le \alpha(\delta)\right)
\le
e^{s\alpha(\delta)} M_{\bar d_j}(-s).
\end{align*}

Let $\varepsilon \in (0,1)$. A sufficient condition for 
$\mathbb{P}(\bar d_j \le \alpha(\delta)) \le \varepsilon$
is
\begin{align*}
e^{s\alpha(\delta)} M_{\bar d_j}(-s) \le \varepsilon,
\end{align*}
which is equivalent to
\begin{align*}
s\alpha(\delta) + \ln M_{\bar d_j}(-s) \le \ln \varepsilon .
\end{align*}

Minimizing over $s > 0$ yields the tightest exponential bound
\begin{align*}
\inf_{s > 0}
\left[
s\alpha(\delta) + \ln M_{\bar d_j}(-s)
\right]
\le \ln \varepsilon .
\end{align*}

Finally, the smallest $\delta \ge 0$ satisfying this condition defines the risk measure in~\eqref{eqn:evar}. \hfill $\square$

\subsubsection{Proof of Proposition \ref{prop:evar_dual}}

The dual interpretation of EVaR follows from the Donsker--Varadhan variational formula. 
For any random variable $X$,
\[
\ln \mathbb{E}_{\mathbb P}[e^{X}]
=
\sup_{Q \ll P}
\left\{
\mathbb{E}_Q[X] - D_{\mathrm{KL}}(Q\|P)
\right\},
\]
where $D_{\mathrm{KL}}(\cdot\|\cdot)$ denotes the Kullback--Leibler divergence.

Applying this identity with $X=-s\bar d_j$ yields
\[
\ln \mathbb{E}_{\mathbb P}[e^{-s\bar d_j}]
=
\sup_{Q\ll P}
\left\{
- s\,\mathbb E_Q[\bar d_j] - D_{\mathrm{KL}}(Q\|P)
\right\}.
\]

Substituting this expression into the EVaR condition gives
\[
\inf_{s>0}
\left[
s\alpha(\delta)
+
\sup_{Q\ll P}
\left\{
- s\,\mathbb E_Q[\bar d_j] - D_{\mathrm{KL}}(Q\|P)
\right\}
\right]
\le \ln \varepsilon .
\]

Dividing by $s>0$ and rearranging yields
\[
\alpha(\delta) +
\inf_{s>0}
\left[
\sup_{Q\ll P}
\left\{
\mathbb E_Q[-\bar d_j]
\right\}
-
s^{-1}\big(D_{\mathrm{KL}}(Q\|P)+\ln\varepsilon\big)
\right]
\le 0 .
\]

This leads to the following dual characterization
\[
\mathrm{E}_\varepsilon(\bar d_j)
=
\inf
\left\{
\delta \ge 0 \;\middle|\;
\alpha(\delta) +
\sup_{Q\ll P}
\left\{
\mathbb E_Q[-\bar d_j]
:\,
D_{\mathrm{KL}}(Q\|P)\le -\ln\varepsilon
\right\}
\le 0
\right\}.
\]

Define
\[
\Psi_\varepsilon(\bar d_j)
:=
\sup_{Q\ll P}
\left\{
\mathbb E_Q[-\bar d_j]
:\,
D_{\mathrm{KL}}(Q\|P)\le -\ln\varepsilon
\right\}.
\]

\[
\mathrm{E}_\varepsilon(\bar d_j)
=
\inf
\left\{
\delta \ge 0
\;\middle|\;
\alpha(\delta) + \Psi_\varepsilon(\bar d_j) \le 0
\right\}.
\]

\hfill $\square$

\subsubsection{Proof of Lemma~\ref{lem:mgf_d_bar_j}}

The result follows from the standard expression for the moment generating function of a Gaussian random variable; see \cite{Kobayashi2011}. \hfill $\square$

\subsubsection{Proof of Theorem \ref{thm:evar_inter_vehicle}}

Using the definition of Entropic Value-at-Risk in Lemma \ref{lem:evar_def} and the MGF from Lemma \ref{lem:mgf_d_bar_j}, we have
\begin{align*}
    E_\varepsilon(\bar{d}_j) 
    &= \inf_{\delta > 0} \left\{ \delta \;\middle|\; 
    \inf_{s > 0} \Big[ s \, \alpha(\delta) - s d + \frac{s^2 \sigma_j^2}{2} \Big] \le \ln \varepsilon
    \right\}.
\end{align*}

\noindent
\textit{Solving the inner minimization over $s$:}
\[
\inf_{s > 0} \Big[ s \, \alpha(\delta) - s d + \frac{s^2 \sigma_j^2}{2} \Big] 
\quad \text{gives} \quad 
s^* = \frac{d - \alpha(\delta)}{\sigma_j^2}.
\]

\noindent
Substituting $s^*$ back into the expression yields
\begin{align*}
\frac{(\alpha(\delta) - d)^2}{2\sigma_j^2} \le - \ln \varepsilon
\quad \implies \quad
\alpha(\delta) - d \le \sqrt{2} \, \sigma_j \sqrt{- \ln \varepsilon}.
\end{align*}

\noindent
Recalling that $\alpha(\delta) = d/(\delta + c)$, we solve for $\delta$:
\[
\delta \ge \frac{1}{1 - \frac{\sqrt{2}\sigma_j}{d} \sqrt{- \ln \varepsilon}} - c.
\]

\noindent
Finally, taking the infimum over $\delta$ gives the desired EVaR. The different branches in \eqref{eqn:evar_inter_vehicle_collision} correspond to the limiting cases where the risk becomes zero (upper branch) or infinite (lower branch) due to the denominator approaching zero.

\hfill$\square$

\subsubsection{Proof of Theorem \ref{thm:evar_laplacian_bounds}}

The proof follows from the monotonicity of the function $f$. Using \eqref{eq:steady_state_var}, we have
\begin{equation} 
c_1 f(\lambda_n\tau,\beta\tau)
\sum_{k=2}^n
(\tilde{\bm e}_j^T q_k)^2 \leq \sigma_j^2
\leq 
c_1 f(\lambda_2\tau,\beta\tau)
\sum_{k=2}^n
(\tilde{\bm e}_j^T q_k)^2,
\end{equation}
where $c_1 = g^2 \frac{\tau^3}{2 \pi}$.

Noting that
\[
\sum_{k=2}^n (\tilde{\bm e}_j^T q_k)^2 = \|\tilde{\bm e}_j\|_2^2 = 2,
\]
we obtain
\begin{equation} 
\sqrt{2 c_1 f(\lambda_n\tau,\beta\tau)}
\leq \sigma_j \leq 
\sqrt{2 c_1 f(\lambda_2\tau,\beta\tau)}.
\end{equation}

Since $E_\varepsilon(\bar{d}_j)$ is decreasing with $\kappa_\varepsilon = d_\varepsilon / (\sqrt{2}\, \sigma_j)$, and $\kappa_\varepsilon$ is decreasing with $\sigma_j$, it follows that $E_\varepsilon(\bar{d}_j)$ is \emph{increasing} with $\sigma_j$. Substituting the upper and lower bounds of $\sigma_j$ into Theorem \ref{thm:evar_inter_vehicle} immediately yields the stated minimum and maximum EVaR bounds.  

\hfill$\square$

% \subsubsection{Analysis of Dual Form of Evar}
% For a random variable $X$, 
% \[\ln \mathbb{E}_\mathbb{P}[e^X] = \sup_{Q \ll P} \{\mathbb{E}_Q[X] - D_{\mathrm{KL}}(Q\|P)\}\].
% Then 
% \begin{align*}
% \inf_{s > 0}
% \left[
% s\alpha(\delta) + \sup_{Q \ll P} \{\mathbb{E}_Q[X] - D_{\mathrm{KL}}(Q\|P)\}
% \right]
% &\le \ln \varepsilon ,\\
% \alpha(\delta) + \inf_{s> 0}
% \left[
%  s^{-1}\sup_{Q \ll P} \{ \mathbb{E}_Q[-s\bar{d}_j] - D_{\mathrm{KL}}(Q\|P)\}
% - s^{-1}\ln \varepsilon \right] &\leq 0 ,
% \end{align*}
% \begin{align*}
%     \inf_{s > 0}
% \left[
%  \sup_{Q \ll P} \{ \mathbb{E}_Q[-\bar{d}_j]\} - s^{-1}(D_{\mathrm{KL}}(Q\|P)
% + \ln \varepsilon) \right] &= \\
% & \hspace{-6cm}\left[
%  \sup_{Q \ll P} \{ \mathbb{E}_Q[-\bar{d}_j]\} - s^{-1}(D_{\mathrm{KL}}(Q\|P)
% + \ln \varepsilon) \right]
% \end{align*}

% Finally 

% \[
% \mathrm{E}_\varepsilon(\bar{d}_j)
% = \inf \{\delta \geq 0 \mid \alpha(\delta) + 
% \sup_{Q\ll P}
% \left\{
% \mathbb E_Q[-\bar{d}_j]
% \;\middle|\;
% D_{\mathrm{KL}}(Q\|P)\le -\ln(\varepsilon)
% \right\}\},
% \]

%%%%%%%%%%%%%%%%%%%%%%%%%%%%%%%%%%%%%%%%%%%%%%%%%%%%%%%%%%%%%%%%%%%%%%%%%%%%%%%%%%%%%%%%%%%%%%
%END OF THE MAIN DOCUMENT
%%%%%%%%%%%%%%%%%%%%%%%%%%%%%%%%%%%%%%%%%%%%%%%%%%%%%%%%%%%%%%%%%%%%%%%%%%%%%%%%%%%%%%%%%%%%%%

% \printbibliography
\end{document}